# The Feshbach shape resonance for high $T_c$ superconductivity in superlattices of nanotubes

Antonio Bianconi

*Dipartimento di Fisica, Università di Roma "La Sapienza", P. le Aldo Moro 2, 00185 Roma, Italy*

The case of a Feshbach shape resonance in the pairing mechanism for high $T_c$ superconductivity in a crystalline lattice of doped metallic nanotubes is described. The superlattice of doped metallic nanotubes provides a superconductor with a strongly asymmetric gap. The disparity and different spatial locations of the wave functions of electrons in different subbands at the Fermi level should suppress the single electron impurity interband scattering giving multiband superconductivity in the clean limit. The Feshbach resonances will arise from the component single-particle wave functions out of which the electron pair wave function is constructed: pairs of wave functions which are time inverse of each other. The Feshbach shape resonance increases the critical temperature by tuning the chemical potential at the Lifshitz electronic topological transition (ETT) where the Fermi surface of one of the bands changes from the one dimensional (1D) to the two dimensional (2D) topology (1D/2D ETT).

## 1 Introduction

While most of the proposed mechanisms for high $T_c$ superconductivity have focused on exotic pairing mechanisms in an effective single band model other authors have proposed the pairing mechanism in a two-component (or multiband) scenario [1]. In the theory of two band superconductivity [2] the condensate many body wave function is determined by the configuration interaction of pairs of opposite spin and momentum in the *a*-band and *b*-band, that on basis of the Bogoliubov transformations is given by

$$|\Psi\rangle = \prod_k (u_k + v_k a^+_{k\uparrow} a^+_{-k\downarrow}) \prod_{k'} (x_{k'} - y_{k'} b^+_{k'\uparrow} b^+_{-k'\downarrow})|0\rangle, \tag{1}$$

where $a^+$ and $b^+$ are creation operators of electrons in the *a* and *b* band respectively and $|0\rangle$ is the vacuum state that includes the configuration interaction between the different pairing channels in *a* and *b* band. The coupling matrix is made of *diagonal* elements describing the *intraband* pairing and *off-diagonal* elements describing the *interband* pairing. The interband pairing terms due to the transfer of a pair from the *a* band to the *b* band and vice-versa is given by

$$\sum_{k,k'} J_{ab}(k,k') \left( a^+_{k\uparrow} a^+_{-k\downarrow} b_{-k'\downarrow} b_{k'\uparrow} \right) \tag{2}$$

where $J_{ab}(k,k')$ is an exchange-like integral [2]. The interband coupling term is not a standard Cooper pairing mechanism in fact it gives an increase of the critical temperature both if it is repulsive or attractive. Moreover it is not a density-density interaction therefore increasing this term does not induce a charge-density wave (CDW) or spin density wave (SDW) instability which compete and suppress the superconducting condensate. Where the exchange interband pairing becomes relevant the effective Cou-lomb repulsion in the pairing goes to zero (superscreening) and the isotope coefficient vanishes also if the phonon mechanism controls the intraband pairing.

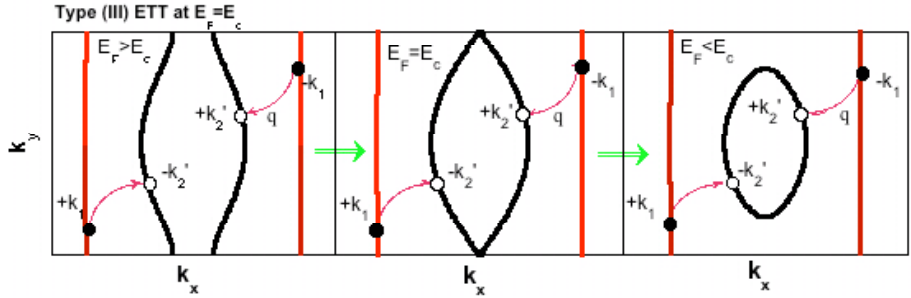

Fig. 1. The Fermi surface of a first and second subband of a superlattice of nanotubes near the 2D to 1D ETT. Going from the left panel to the right panel the chemical potential $E_F$ crosses a vHs singularity at $E_c$ associated with the change of the 1D Fermi topology of the second subband for $E_F>E_c$ to a 2D topology for $E_F<E_c$, while the Fermi surface of the first subband retains its one-dimensional (1D) character. A relevant interband pairing process for the Feshbach shape resonance with the transfer of a pair from the first to the second subband and viceversa is shown.

Multi-band or two-band superconductivity occurs only in the "clean limit" where the single electron interband impurity scattering rate $\gamma_{ab} \ll (K_B/\hbar)T_c$. Nearly all superconductors (with few exceptions [1]) are in the "dirty limit" [3] where the impurity scattering mixes the electron wave functions on different spots and different bands on the Fermi surfaces reducing the system to an effective BCS single band system. In the *clean limit* the anisotropic gap parameter $\Delta_k$ depends on the directions of the wave-vector k with respect to the crystal lattice and on the different bands. The k-dependent gap $\Delta_k$ and $T_c$ depend on the details of the electronic structure, i.e. on the electron wave functions near the Fermi level that form the pair wave functions. On the contrary in the *dirty limit* the gap k-dependence is averaged out by impurity scattering and the gap $\Delta$ depends on the electronic structure mainly via the total density of states. The electronic structure of the material architecture made of superlattices of metallic units: dots, or wires, or layers [4-7] at atomic limit can favor the increase of $T_c$ in the clean limit. First, superlattices is predicted to show multiband superconductivity in the clean limit since the interband scattering rate $\gamma_{ab}$ between particular subbands with different parity and different spatial distribution is strongly suppressed. Second, in these materials by tuning the chemical potential near a Lifshitz electronic topological transition (ETT) where the Fermi surface topology of one of the subbands changes its dimensionality, as it is shown in Fig.1, the critical temperature shows a *Feshbach shape resonance* [1,4-7].

The theory of the *shape resonances* observed in neutron scattering cross section in nuclear physics was developed by Herman Feshbach as due to the configuration interaction between different scattering channels [8] first introduced by Ugo Fano for the two electron excitations in the absorption spectrum of helium [9]. It has been used recently to achieve the Bose-Einstein condensation (BEC) in the ultracold dilute bosonic gas of alkali atoms [10] and recently to get a BCS condensate in a fermionic gas [11]. The theory of shape resonance in multiband superconductivity was first presented by Blatt and Thompson for a thin metallic membrane [12] where the Fermi level is tuned in the proximity of the bottom or the top of a band where a new detached Fermi surface appears or disappears. It was developed for *heterostucture at the atomic limit* made of superlattices in Rome by our group [1,4-7]. Recently it has been found that the record for the highest $T_c$ (40K) in intermetallics is reached in a *heterostructure at the atomic limit*, i,e., a superlattice of graphene-like boron layers intercalated by magnesium ($MgB_2$) forming a superconducting phase in the clean limit where $T_c$ is enhanced by tuning the chemical potential in the proximity of the 2D to 3D ETT in the σ band [13,1].

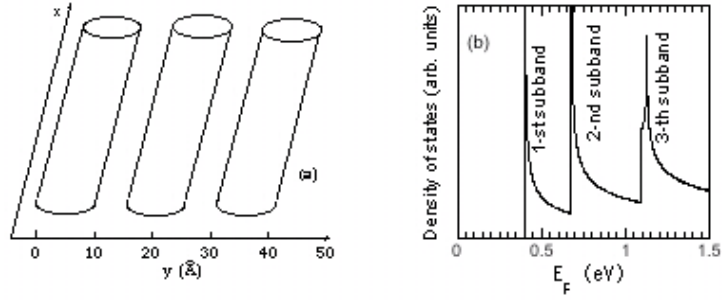

Fig. 2. *Panel a:* The pictorial view of a 2D superlattice of nanotubes with a period $\lambda_p$ =1.55 nm in the y direction
*Panel b :* the density of states of the superlattice of nanotube including the electron hopping between the nanotubes.

## 2. High $T_c$ in a superlattice of nanotubes

Graphite and fullerene [14,15] crystals provide the very simple case of a superlattice of monolayers (graphene layers) and dots (buckyballs) respectively. Recently the feasibility of growing 3D ordered architectures of carbon nanotubes has been shown [16]. The chemical potential and the electronic properties of these systems can be changed by doping with electrons or holes by chemical intercalation. This approach has been used extensively for compounds such graphite intercalated compounds (GIC) [14] and fullerenes (FIC) [15]. Alkali metal intercalation of mats of bundled single wall carbon nanotubes (SWCNT) takes place inside the channels of the triangular bundle lattice [17] and leads to a shift of the Fermi energy, a loss of the optical transitions and an increase of the conductivity by about a factor of thirty [18]. A complete charge transfer between the donors and the SWCNT was observed up to saturation doping, which was achieved at a carbon to alkali metal ratio of about seven [18]. The change of the low energy electronic properties in potassium intercalated bundles of single wall carbon nanotubes as a function of doping using photoemission has revealed a Fermi edge and the feasibility of tuning the chemical potential up to reach the third subband at high doping [19].

In this work we consider a crystal made of a *metal heterostuctures at the atomic limit:* a superlattice of nanotubes where the Fermi level is tuned by electron doping. We show that high $T_c$ superconductivity by shape resonance is possible by tuning the chemical potential to the 2D/1D ETT in the second or third subband. Let us consider a 2D superlattice of carbon nanotubes of period $\lambda_p$ on a x,y plane shown in Fig. 2. The system is modeled with free electrons moving in the nanotube direction x. The charge carriers have to overcome a periodic potential barrier V(x,y), with period $\lambda_p$, amplitude $V_b$ and width W (the separation between the nanotubes) along the y direction, determined by the potential barrier between the nanotubes.

The wave-functions solution of the Schrödinger equation are

$$\psi_{n,k_x,k_y}(x,y) = e^{ik_x x} \cdot e^{ik_y q \lambda_p} \psi_{n,k_y}(y) \qquad (3)$$

where in the nanotube

$$\psi_{n,k_y}(y) = \alpha e^{ik_w \tilde{y}} + \beta e^{-ik_w \tilde{y}} \ for |\tilde{y}| < L/2; \ k_w = \sqrt{2m_w\left(E_n(k_y) + V_b\right)/\hbar^2} \qquad (4)$$

and in the barrier

$$\psi_{n,k_y}(y) = \gamma e^{ik_b\tilde{y}} + \delta e^{-ik_b\tilde{y}} \quad for |\tilde{y}| \geq L/2; \; k_b = \sqrt{2m_b E_n(k_y)/\hbar^2} \tag{5}$$

The coefficients α, β, γ and δ are obtained by imposing the Bloch conditions with periodicity $\lambda_p$, the continuity conditions of the wave function and its derivative at L/2, and finally by normalization in the surface unit. The solution of the eigenvalue equation for E gives the electronic energy dispersion for the n subbands. There are $N_b$ solutions for $E_n(k_y)$, with $1 \leq n \leq N_b$, for each $k_y$ in the Brillouin zone of the superlattice giving a dispersion in the y direction of the $N_b$ subbands with $k_x=0$.

The partial density of states (DOS) of the n-th subband gives a step-like increase of the total DOS when the chemical potential reaches the bottom of the subband n=1,2,3 where a type (I) ETT occurs as shown in Fig. 2. The DOS peaks in Fig. 2 are due to partial DOS of each subband tuning the chemical potential by electron doping at the vHs in each subband. The panel (a) of Fig. 3 shows the details of the DOS near the type (III) ETT at $E_c$ in the third subband as a function of the reduced Liftshitz parameter measuring the distance from the electronic topological transition at $E=E_c$, $\varsigma = (E_F - E_c)/D$ where D is the dispersion of the third subband in the y direction of the superlattice, transversal to the nanotube direction.

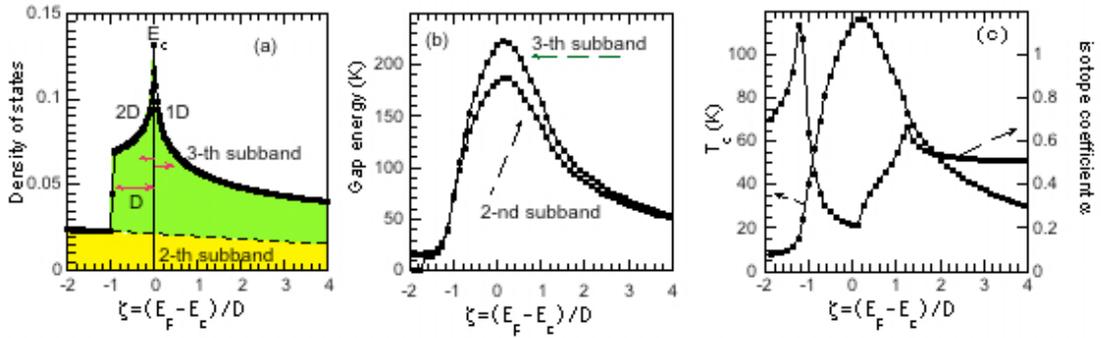

*Fig 3.* Panel (a) the DOS near the bottom of the third subband of the superlattice of nanotubes as function of the reduced Liftshitz parameter ζ where D=36.6 meV is the dispersion of the third subband in the y direction of the superlattice, The 2D to 1D ETT occurs at ζ=0. Panel (b): The superconducting gaps in the second, $\Delta_2$, and third, $\Delta_3$, subband. Panel (b): The critical temperature $T_c$ and the isotope coefficient α at the shape resonance.

The Fermi surface of the third subband changes from the 1D topology to 2D tuning the chemical potential going from $E_F > E_c$ to $E_F < E_c$ respectively, i.e., crossing ζ=0, as it is shown in Fig. 1.
The superlattice with its characteristic wavevector $q=2\pi/\lambda_p$ induces a relevant <u>k dependent</u> interband pairing interaction $V_{n,n}(k,k')$. This is the non BCS interband effective pairing interaction (of any repulsive or attractive nature) with a generic cutoff energy $\hbar\omega_o$. The interband interaction is controlled by the details of the *quantum superposition of states corresponding to different spatial locations i.e,* between the wave functions of the pairing electrons in the different subbands of the superlattice

$$V_{n,n}(k,k') = V^o_{n,k_y;n',k_y'}\theta(\hbar\omega_0 - |\varepsilon_n(k) - \mu|)\,\theta(\hbar\omega_0 - |\varepsilon_n(k') - \mu|) \text{ where } k = (k_x, k_y) \text{ and}$$

$$V^o_{n,k_y;n',k'_y} = -J \int_S dx\,dy\, \psi_{n,-k}(x,y)\psi_{n',-k'}(x,y)\psi_{n,k}(x,y)\psi_{n',k'}(x,y)$$

$$= -J \int_S dx\,dy\, |\psi_{n,k}(x,y)|^2 |\psi_{n',k'}(x,y)|^2 \quad (6)$$

n and n' are the subband indexes. $k_x$ ($k_x'$) is the component of the wavevector in the wire direction (or longitudinal direction) and $k_y$, ($k_y'$) is the superlattice wavevector (in the transverse direction) of the initial (final) state in the pairing process, and μ is the chemical potential.

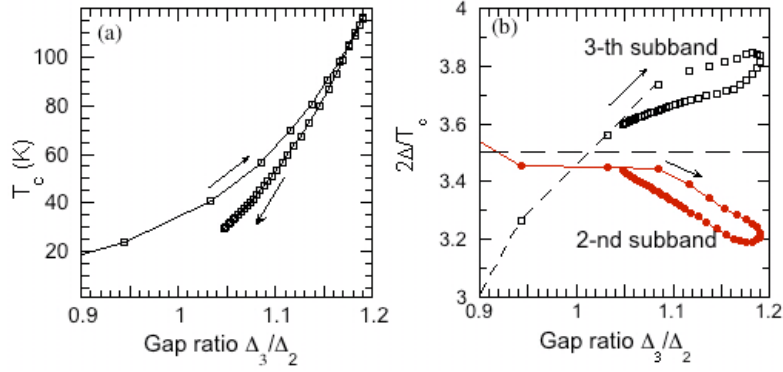

*Fig.* 4. Panel (a) The variation of $T_c$ as a function of the gap ratio $R = \Delta_3/\Delta_2$ at the shape resonance. The maximum $T_c$ is reached for the maximum gap ratio. Panel (b) the ratio $2\Delta_2/T_c$ and $2\Delta_3/T_c$ as a function of the gap ratio $\Delta_3/\Delta_2$. The largest deviation from the single band BCS value $2\Delta/T_c = 0.35$ is reached at the highest gap ratio indicating the key effect of multiband superconductivity.

In the separable kernel approximation, the gap parameter has the same energy cut off $\hbar\omega_o$ as the interaction. Therefore it takes the values $\Delta_n$ ($k_y$) around the Fermi surface in a range $\hbar\omega_o$ depending from the subband index and the superlattice wavevector $k_y$.

The self consistent equation, for the ground state energy gap $\Delta_n$ ($k_y$) is:

$$\Delta_n(\mu,k_y) = -\frac{1}{2N} \sum_{n',k'_y,k'_x} \frac{V_{n,n'}(k,k') \cdot \Delta_{n'}(k'_y)}{\sqrt{(E_{n'}(k'_y) + \varepsilon_{k'_x} - \mu)^2 + \Delta^2_{n'}(k'_y)}} \quad (7)$$

where N is the total number of wavevectors. Solving iteratively this equation gives the anisotropic gaps dependent on the subband index and weakly dependent on the superlattice wavevector $k_y$. The structure in the interaction gives different values for the gaps $\Delta_n$ giving a system with an anisotropic gaps in the different segments of the Fermi surface.

The superconducting gaps in the second, $\Delta_2$, and third, $\Delta_3$, subband in a superlattice of nanotubes as a function of the reduced Lifshitz parameter ζ are shown in Fig. 3b, where $E_F$ is tuned at $E_c$ for ζ =0, the energy cut off for the pairing interaction is fixed at 500K. The increase of the gap $\Delta_2$ is driven only by the Feshbach resonance in the interband pairing since the partial DOS of the second subband has not peaks.

The critical temperature $T_c$ of the superconducting transition can be calculated by iterative method

$$\Delta_n(k) = -\frac{1}{N}\sum_{n'k'} V_{nn'}(k,k') \frac{\tgh(\frac{\xi_{n'}(k')}{2T_c})}{2\xi_{n'}(k')} \Delta_{n'}(k') \tag{8}$$

where $\xi_n(k) = \varepsilon_n(k) - \mu$.

The interband pairing term enhances $T_c$, as shown in Fig. 3c, by tuning the chemical potential in an energy window around the Van Hove singularity, $\zeta=0$, associated with a change of the topology of the Fermi surface from 1D to 2D of one of the subbands of the superlattice in the clean limit.
The the isotope coefficient $\alpha$ and the critical temperature $T_c$ at the shape resonance are shown in Fig. 3c. The result shows the characteristic feature of the $T_c$ amplification by a shape resonance at the maximum critical temperature $T_{c,max}$ the isotope coefficient is close to zero or negative, and shows large positive values up to 1.2 much larger that the standard BCS value $\alpha=0.5$ at the bottom of the band $\zeta=-1$.

Fig. 4a shows the variation of $T_c$ as a function of the gap ratio $\Delta_3/\Delta_2$ showing that the critical temperature increases by increasing the gap ratio. This provides direct evidence that is a measure of the relevance of the shape resonance in interband pairing. Fig. 4b shows the ratio $2\Delta_2/T_c$ and $2\Delta_3/T_c$ as a function of the gap ratio. It shows large deviations from the BCS value $2\Delta/T_c = 0.35$ for a single band indicating the key effect of multiband superconductivity.

In conclusion we have shown that doped single crystals of nanotubes provide *metal heterostuctures at the atomic limit* that show high $T_c$ superconductivity driven by the shape resonance by tuning the chemical potential at a dimensional electronic topological transition.

**Acknowledgements** This work is supported by MIUR in the frame of the project Cofin 2003 " Leghe e composti intermetallici: stabilità termodinamica, proprietà fisiche e reattività" on the "synthesis and properties of new borides"